\documentclass[3p,times,procedia]{elsarticle}
\usepackage{nupha_ecrc}
\pdfoutput=1

\volume{00}

\firstpage{1}

\journalname{Nuclear Physics A}

\runauth{}

\jid{nupha}

\jnltitlelogo{Nuclear Physics A}

\usepackage{amssymb}
\usepackage[figuresright]{rotating}

\newcommand{\pt}{\ensuremath{p_{\mathrm{T}}}}

\newcommand{\mubinv}{\mathrm{\mu b^{-1}}}
\newcommand{\pbinv}{\mathrm{pb^{-1}}}
\newcommand{\sqrtsNN}{\ensuremath{\sqrt{\mathrm{s_{NN}}}}}
\newcommand{\Njets}{\ensuremath{N_{\mathrm{jets}}}}
\begin{document}

\begin{frontmatter}

%% Title, authors and addresses

%% use the tnoteref command within \title for footnotes;
%% use the tnotetext command for the associated footnote;
%% use the fnref command within \author or \address for footnotes;
%% use the fntext command for the associated footnote;
%% use the corref command within \author for corresponding author footnotes;
%% use the cortext command for the associated footnote;
%% use the ead command for the email address,
%% and the form \ead[url] for the home page:
%%
%% \title{Title\tnoteref{label1}}
%% \tnotetext[label1]{}
%% \author{Name\corref{cor1}\fnref{label2}}
%% \ead{email address}
%% \ead[url]{home page}
%% \fntext[label2]{}
%% \cortext[cor1]{}
%% \address{Address\fnref{label3}}
%% \fntext[label3]{}

%% Instructions from Editor: Please use the following \dochead only in the preprint version (e-print arXiv etc.); 
%% use empty \dochead{} when submitting to Nuclear Physics A!
\dochead{XXVIIIth International Conference on Ultrarelativistic Nucleus-Nucleus Collisions\\ (Quark Matter 2019)}
%\dochead{}
%% Use \dochead if there is an article header, e.g. \dochead{Short communication}
%% \dochead can also be used to include a conference title, if directed by the editors
%% e.g. \dochead{17th International Conference on Dynamical Processes in Excited States of Solids}

\title{Studies of Quark and Gluon Contributions to Jets using Jet Charge Measurements in pp and PbPb Collisions}

%% use optional labels to link authors explicitly to addresses:
%% \author[label1,label2]{<author name>}
%% \address[label1]{<address>}
%% \address[label2]{<address>}

\author{Dhanush Anil Hangal on behalf of the CMS Collaboration}

\address{University of Illinois at Chicago}
%\ead{dhanush.anil.hangal@cern.ch}

\begin{abstract}
Jet charge, defined as the momentum-weighted sum of the electric charges of particles inside a jet is sensitive to the electric charge of the initiating parton and can be used to study the color charge dependence of the parton energy loss in the QGP. In this paper, the first measurements of jet charge in heavy-ion collisions are presented using lead-lead (PbPb) collision data and compared to results from proton-proton (pp) data at the same collision energy. The measurements are unfolded for detector and background effects and are studied differentially in \pt~ and additionally as a function of collision centrality in PbPb collisions. We also present a template fitting technique for estimating the fractions of quark- and gluon-initiated jets in pp and PbPb collisions based on Monte Carlo templates. This analysis uses pp and PbPb collision data collected by the CMS experiment at $\sqrt{\mathrm{s_{NN}}}=5.02$ TeV.
\end{abstract}

\begin{keyword}

CMS, Physics, Computing 
%% keywords here, in the form: keyword \sep keyword

%% MSC codes here, in the form: \MSC code \sep code
%% or \MSC[2008] code \sep code (2000 is the default)

\end{keyword}

\end{frontmatter}

%%
%% Start line numbering here if you want
%%
%\linenumbers

%% main text
\section{Introduction}
\label{introduction}

A deconfined state of quarks and gluons called the quark-gluon plasma (QGP) is predicted to exist in extremely high temperature and density conditions~\cite{Karsch_1995}. High transverse momentum (\pt) jets, originating from partons produced in the initial hard scatterings in relativistic heavy-ion collisions, are expected to suffer energy loss in the QGP medium. This phenomenon known, as ``jet quenching'', has been studied extensively via measurements at the CERN LHC~\cite{Chatrchyan:2011sx}. The energy loss mechanisms in the QGP, including their color dependence, are still not fully understood~\cite{Majumder-2011}.

Several recent studies have interpreted results based on a modification in the fractions of quark- and gluon-initiated jets in PbPb collisions compared to the pp baseline due to the color charge dependence of jet quenching in the QGP~\cite{Spousta2016}. The momentum-weighted sum of the electric charges of particles inside a jet, known as the ``jet charge'', is expected to be sensitive to the electric charge of the initiating parton~\cite{Feynman:1978}. Based on the fundamental difference in the electric charge of quarks and gluons, a template-fitting method is employed to extract the respective jet fractions in pp and PbPb collisions, using templates from Monte Carlo simulations. Although jet charge is not infrared-safe, the dependence of the mean and width (standard deviation) of the jet charge distribution on both jet energy and jet cone size, can be calculated independently of MC fragmentation models~\cite{krohn2013}. This makes jet charge a reliable variable for the template fitting procedure.

Jet charge measurements are presented using pp and PbPb collision data collected with the CMS detector~\cite{Collaboration_2008} at $\sqrtsNN = 5.02$ TeV~\cite{CMS-PAS-HIN-18-018}. An unfolding procedure is used to correct the measurements for detector and background effects. The results are presented in bins of the minimum \pt~ threshold of particles used in the measurement and also as a function of the overlap of the two colliding Pb nuclei (centrality), with fully-overlapping nuclei defined as ``0\% central''.

\section{Analysis and Results}
\label{analysis}

Data samples of PbPb and pp collisions collected at the CMS detector at $\sqrtsNN = 5.02$ TeV, corresponding to integrated luminosities of 404 $\mubinv$ and 27.4 $\pbinv$, respectively, are used in these measurements. Events are selected using a jet trigger requiring at least one jet with $\pt>100$ GeV. \textsc{PYTHIA} event generator (version 6.424~\cite{Sjostrand2014zea}, tune Z2) is used to produce the hard-scattering MC samples (referred to as \textsc{PYTHIA6}). For PbPb studies, the \textsc{PYTHIA6} interactions are embedded into simulated minimum-bias PbPb events produced with \textsc{HYDJET} (version 1.383~\cite{Lokhtin:2005px}). Jets are reconstructed using the anti-kT algorithm~\cite{bib_antikt} with a distance parameter R = 0.4, and the contributions of the underlying event to the jet \pt~ are subtracted using the “noise/pedestal subtraction“ technique in in PbPb collisions~\cite{Kodolova:2007hd}. Reconstructed charged tracks within $|\eta| < 2.4$ and having $\pt>1$ GeV are used in the jet charge measurements. 

The jet charge is defined as: 

\begin{equation} \label{eq:jet_charge_eq}
Q^\kappa = \frac{1}{({\ensuremath{p_{\mathrm{T,jet}}}})^\kappa} \sum_{i\in \mathrm{jet}} q_{i} \ensuremath{p_{\mathrm{T},i}}^\kappa~,
\end{equation}

where {\ensuremath{p_{\mathrm{T,jet}}}} represents the transverse momentum of the jet, and $q_i$ and \pt$_{,i}$ refer to the electric charge (in terms of the proton charge $\mathrm{e}$) and transverse momentum of the i--th particle in the jet cone, respectively. The $\kappa$ parameter controls the sensitivity of the jet charge to the particle \pt~ in the jet cone~\cite{Feynman:1978}. High (low) values of $\kappa$ enhance the contribution from high-(low-)\pt~ particles to the measured jet charge. Theoretical calculations indicate that $\kappa$~$\approx$~0.5 is the most sensitive to the electric charge of the initiating parton in vacuum~\cite{waalewijn2012}. In this work, jet charge measurements are shown for $\kappa$ values of 0.3, 0.5, and 0.7, and with different selections on the minimum track \pt~ of 1, 2, 4, and 5 GeV. The jet charge measurements at the detector-level are affected by track reconstruction inefficiencies, and additionally by background in PbPb collisions. For a track \pt~ selection of 2 GeV and $\kappa$ value of $0.5$, going from the particle- to the detector-level, these effects lead to an increase in the width of jet charge distributions by 7\% in pp collisions and up to 23\% in the most central PbPb collisions. D'Agostini iterative method~\cite{DAgostini1995487}, as implemented in the RooUnfold software package~\cite{Adye:2011gm}, is used to unfold the jet charge distributions for these effects. A data-driven method is employed to study the MC bias and corresponding systematic uncertainties due to background unfolding in PbPb measurements. This is performed by measuring jet charge using jets in a jet-triggered event and tracks within the jet cone from a matched minimum-bias event. The jet-triggered and minimum-bias events are matched according to their primary vertex and collision centrality values. The background contribution to jet charge  obtained from this technique is observed to be within $1\%$ agreement with that from  HYDJET. 

To extract the fractions of quark- and gluon-initiated jets in data, jet charge distributions of different flavor jets in MC samples are used as templates to fit the unfolded data jet charge distributions. In the \pt~ range examined in this analysis, up quark-, down quark- and gluon-initiated jets constitute a dominant fraction of the sample, and are thus used as the nominal templates in the fitting procedure. The up antiquark ($\mathrm{\overline{u}}$) and down antiquark ($\mathrm{\overline{d}}$) jet fractions are varied along with the up and down quark jets, respectively. The fraction of jets initiated by charm, strange, and bottom (anti)quarks, referred together as ``other flavor'' jets, is estimated from MC and kept fixed in the nominal fitting procedure. The corresponding systematic uncertainty is estimated by varying the ``other flavor'' jets by their total fraction and repeating the fitting procedure. The resulting deviation from the nominal fitting result is then assigned as a systematic uncertainty. The mean of the jet charge distributions for the individual flavors varies by less than 1\% as a function of the jet \pt, enabling a stable fitting procedure. Additional corrections are applied to account for jets with no reconstructed tracks above the \pt~ threshold and observed calorimeter response biases.

\begin{figure}[t]
\begin{center}
\includegraphics[width=0.99\textwidth]{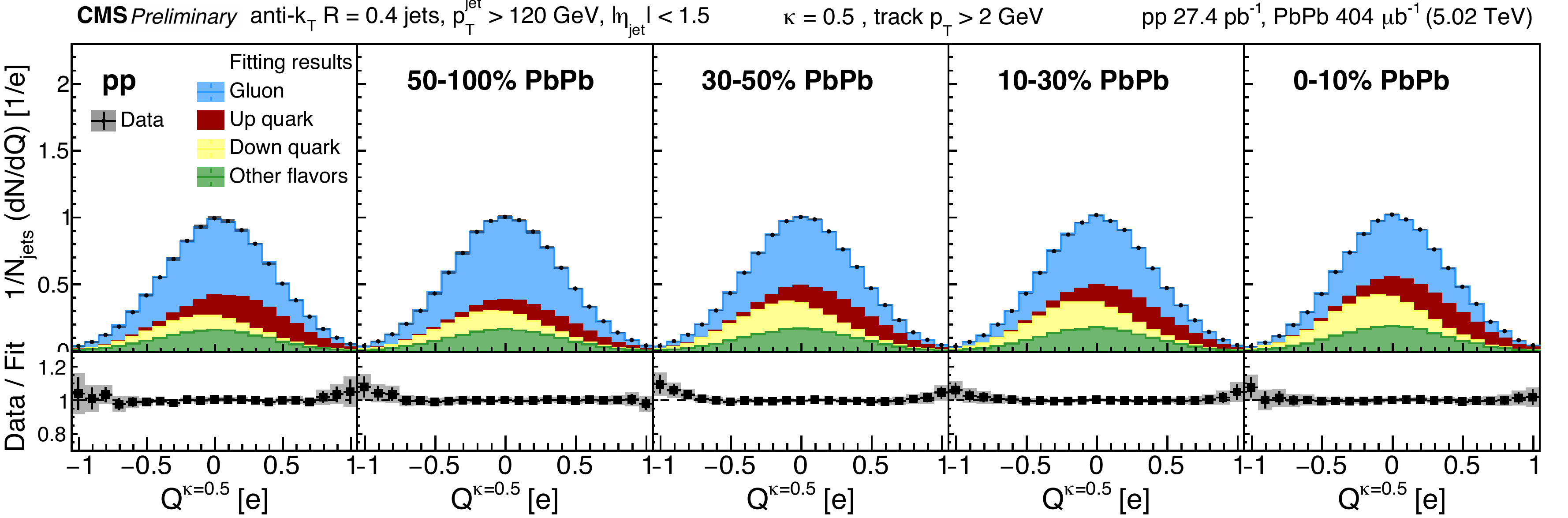}
\caption{(Upper) Jet charge measurements shown for inclusive jets in data along with the extracted fractions of up, and down quark jets, gluon jets, and the ``other flavor'' jets. The systematic and statistical uncertainties in the distributions are shown by the shaded regions and vertical bars, respectively. The jet charge measurements shown here are for $\kappa$ = 0.5 and a minimum track \pt~ of 2 GeV. (Lower) Ratio of the jet charge measurements to the results of template fits. For more details refer to Ref.~\cite{CMS-PAS-HIN-18-018}.}
\label{fig:fitting_ratio}
\end{center}
\end{figure}

The unfolded jet charge measurements, normalized to the total number of jets in the sample \Njets, are shown in the upper panel of Fig.~\ref{fig:fitting_ratio} for a sample selection of $\kappa$ = 0.5 and a minimum track \pt~ of 2 GeV. The measurements are shown with solid black points and the extracted fraction of quark and gluon-initiated jets are displayed as stacked histograms. The lower panel of the figure shows the ratio of the data over the results from the template fits, where no significant deviation from unity is observed in the entire fitting range. 

\begin{figure}[h]
\begin{center}
\includegraphics[width=0.99\textwidth]{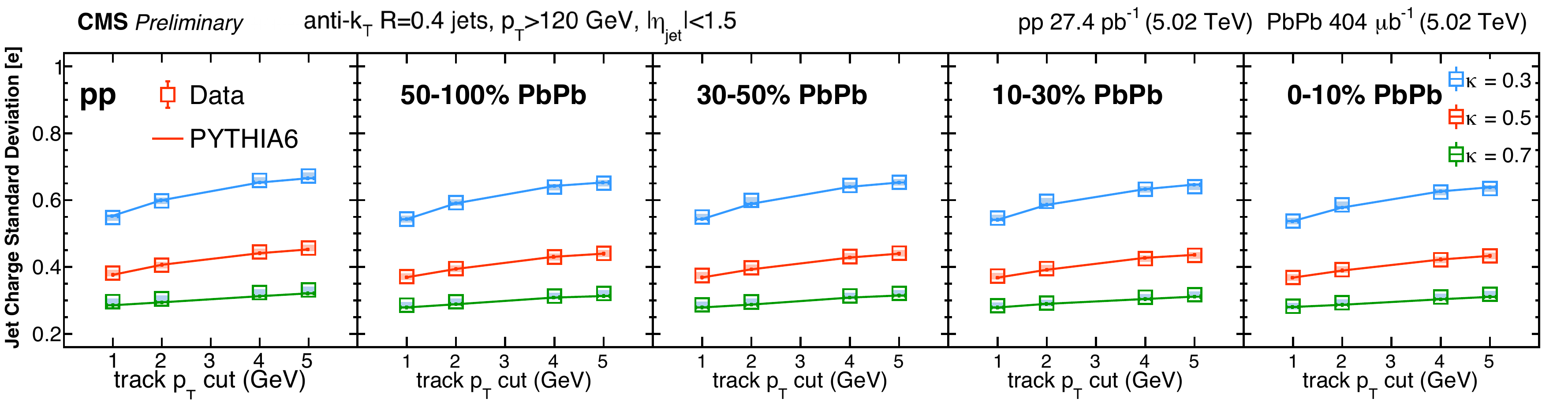}
\caption{The standard deviation of the jet charge distributions with different $\kappa$ values and track \pt~ thresholds for pp and PbPb collisions compared with the \textsc{PYTHIA6} predictions. The systematic and statistical uncertainties in the standard deviation measurements are shown by the shaded regions and vertical bars, respectively. For more details refer to Ref.~\cite{CMS-PAS-HIN-18-018}.}
\label{fig:jetcharge_width}
\end{center}
\end{figure}

Theoretical predictions incorporating color-charge dependence into jet energy loss calculations predict a reduced fraction of gluon-initiated jets in PbPb collisions compared to pp due to stronger quenching of gluon jets~\cite{Li:2019dre}. An increase in the width of the measured jet charge distributions is thus expected as a result of the reduced gluon jet fraction in a quenched sample. From Fig.~\ref{fig:jetcharge_width}, it is observed that the data jet charge width has very little dependence on centrality and is also well described by the MC, where no quenching effects are modeled.

\begin{figure}[h]
\begin{center}
\includegraphics[width=0.99\textwidth]{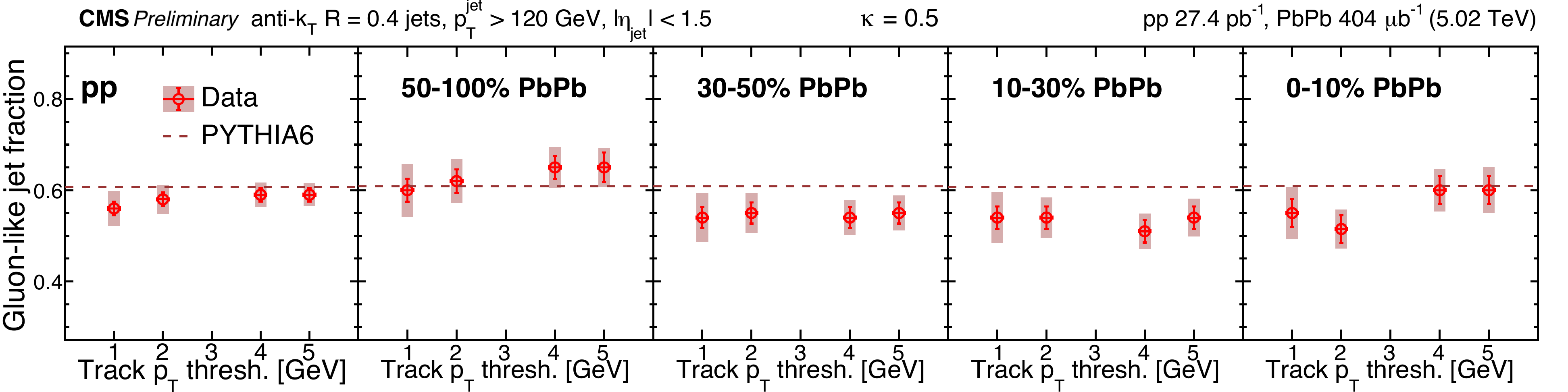}
\caption{Fit results for the extraction of gluon-like jet fractions in pp and PbPb data shown for $\kappa$ = 0.5 and for different track \pt~ threshold values. The systematic and statistical uncertainties are represented by the shaded regions and vertical bars, respectively. The \textsc{PYTHIA6} prediction for the fraction of gluon-initiated jets is shown in dashed red lines~\cite{CMS-PAS-HIN-18-018}.}
\label{fig:fitting_qg_ptcut}
\end{center}
\end{figure}

\begin{figure}[h]
\begin{center}
\includegraphics[width=0.99\textwidth]{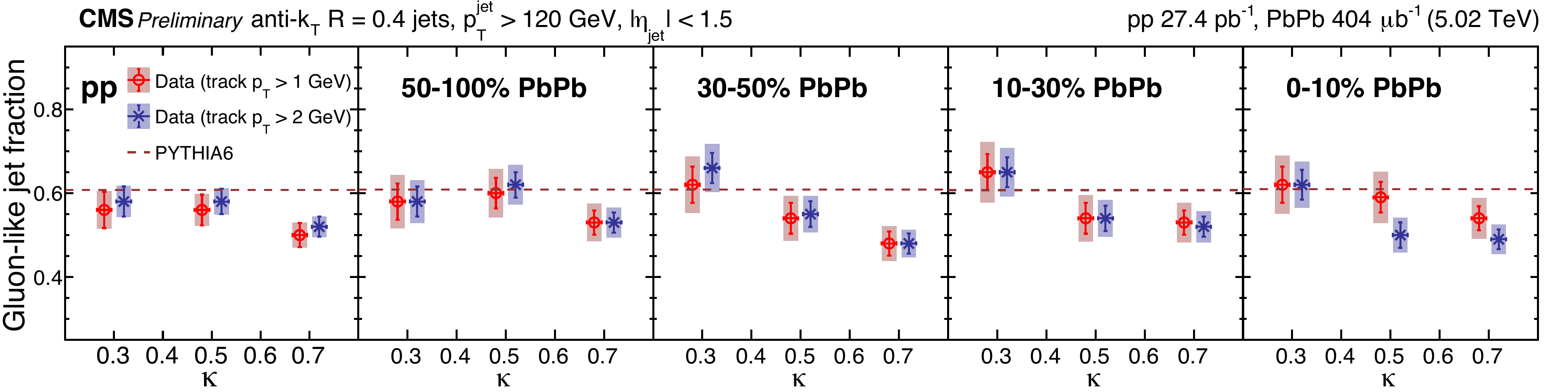}
\caption{Fit results similar to Fig.~\ref{fig:fitting_qg_ptcut}, but shown for different $\kappa$ values, and for track \pt~ thresholds of $>1$, and $>2$ GeV~\cite{CMS-PAS-HIN-18-018}.}
\label{fig:fitting_qg_kappa}
\end{center}
\end{figure}

The results for the extracted fractions of quark- and gluon-like jets in the inclusive sample are shown in Fig.~\ref{fig:fitting_qg_ptcut} and Fig.~\ref{fig:fitting_qg_kappa} as a function of the track \pt~ threshold and $\kappa$, respectively. No significant modification is observed in the relative fractions of the quark- and gluon-like jets in central PbPb collisions compared to peripheral PbPb and pp collisions from these results. 

\section{Summary}
\label{summary}

The first measurements of jet charge in heavy-ion collisions are presented using PbPb data collected with the CMS detector and compared to results from pp data at the same collision energy~\cite{CMS-PAS-HIN-18-018}. The jet charge distribution widths for pp and PbPb collisions are observed to be in good agreement with predictions from \textsc{PYTHIA6}. A template-fitting method is presented to extract the fractions of quark- and gluon-initiated jets in pp and PbPb collisions using templates from MC samples. No significant modification is observed in the extracted gluon-like or quark-like jet fractions in a sample of jets with $\pt>120$ GeV in all studied PbPb centrality bins and also compared to the pp results.

%% The Appendices part is started with the command \appendix;
%% appendix sections are then done as normal sections
%% \appendix

%% \section{}
%% \label{}

%% References
%%
%% Following citation commands can be used in the body text:
%% Usage of \cite is as follows:
%%   \cite{key}         ==>>  [#]
%%   \cite[chap. 2]{key} ==>> [#, chap. 2]
%%

%% References with BibTeX database:

\bibliographystyle{elsarticle-num}
\bibliography{jetcharge}

%% Authors are advised to use a BibTeX database file for their reference list.
%% The provided style file elsarticle-num.bst formats references in the required Procedia style

%% For references without a BibTeX database:

% \begin{thebibliography}{00}

%% \bibitem must have the following form:
%%   \bibitem{key}...
%%

% \bibitem{}

% \end{thebibliography}

\end{document}